\documentclass{appolb}
\usepackage{graphicx}
\usepackage{amsmath} 
\usepackage{amssymb} 

\begin{document}
\title{Prenematic Fluctuations in Nanoparticle-Hosted Systems%
}

\author{Szymon Starzonek
\address{Institute of Theoretical Physics and Mark Kac Center for Complex Systems Research, Jagiellonian University, Krak\'{o}w, Poland}
\\[3mm]
{Krzysztof G\'{o}rny, Zbigniew Dendzik 
\address{Faculty of Science and Technology, University of Silesia in Katowice, Chorz\'{o}w, Poland}
}
\\[3mm]
{Dejvid \v{C}re\v{s}nar 
\address{Institute of High Pressure Physics PAS, Warszawa, Poland}
}
\\[3mm]
Ale\v{s} Igli\v{c}
\address{Laboratory of Physics, Faculty of Electrical Engineering, University of Ljubljana, Ljubljana, Slovenia}
}
\maketitle
\begin{abstract}
This study combines broadband dielectric spectroscopy (BDS) experiments with molecular dynamics (MD) simulations to investigate the influence of nanoparticle (NP) inclusions on pretransitional phenomena in a liquid crystal (LC) host. We aimed to fill the existing gap between macroscopic observations and their microscopic origins. Our experimental results on SiO$_2$-doped 5CB composites demonstrate that while NP additions do not significantly change the isotropic-nematic transition temperature ($T_c$), the pretransitional effects exhibit universal behavior, confirmed by identical critical exponents across all samples. The observed systematic decrease in dielectric permittivity is explained by MD simulations, which reveal that nanoparticles act as "seeds" for topological defects, forcing the surrounding LC molecules into a "hedgehog" configuration. This static, defect-induced structure leads to a local antiparallel alignment and cancellation of molecular dipoles, providing a direct microscopic mechanism for the macroscopic dielectric response and successfully bridging the micro-macro scales.\end{abstract}
  
\section{Introduction}
Liquid crystal (LC) systems doped with nanoparticles (NPs) are attracting growing interest due to their fundamental properties and potential for innovative applications [1-3]. Over the past two decades, an increasing number of theoretical, experimental, and applied studies have explored this topic [2-15, 18-35]. The introduction of nanoparticles into a liquid crystalline host is a promising method for modifying its physical properties—such as electrical and thermal conductivity, switching voltage, or dielectric anisotropy—without the need for new chemical synthesis [4,5,7,10,12, 13,30]. Furthermore, the use of appropriate polymer substrates serving as microcells filled with nanoparticle-doped liquid crystals enables the creation of communication antennas and radiation sensors [28,30].

From a fundamental research perspective, a particularly interesting area is the pretransitional phenomena attending the phase transition from the isotropic phase to a mesophase (e.g., nematic (I-N), smectic (I-Sm), or chiral) [1-3, 17-20, 26,27,29]. While these effects, associated with critical-like prenematic fluctuations, are well-described for pure liquid crystals, the influence of nanoparticles on their character and universality remains incompletely understood. Extensive experimental evidence exists for nematic, smectic, and cholesteric phases doped with nanoparticles [3,6,10,17-20,26,27]. However, it must be emphasized that previous studies of LC+NP composites have sometimes yielded contradictory results regarding phase transition characteristics. In particular, a direct link between macroscopic measurements and the specific, microscopic arrangements of molecules around the nanoparticles has been lacking. Moreover, early theoretical models are not always corroborated by experimental findings [2-5,7,18-20,26,27].

The aim of this work is to bridge this gap through an integrated approach that combines dielectric experiments with molecular dynamics simulations. We focus on verifying the universality of pretransitional phenomena in LC+NP composites and on elucidating the molecular mechanism responsible for the evolution of the dielectric constant near the I-N transition. Specifically, we seek to explain the microscopic origin of the characteristic "bending" behavior observed in the temperature dependence of the dielectric permittivity.

The dielectric method serves as a potent tool for investigating the ordering and reorientation of polar molecules, with various theories explaining such behaviours. For example, the Onsager approach is suitable for describing the isotropic phase in liquid-crystalline materials (above $T_c$), while the analysis of nematic phases often involves the application of theories such as Maier-Saupe, Maier-Meier, or Landau-de Gennes [1,2]. In essence, the Maier-Saupe description can be employed as an equivalent representation.

For this reason, we use Broadband Dielectric Spectroscopy (BDS) for the precise characterization of macroscopic pretransitional anomalies and molecular dynamics [16-20,26,27]. In parallel, using molecular dynamics simulations, we visualize and analyze the local ordering of LC molecules around a nanoparticle. As we demonstrate, the combination of these two methods allows for the construction of a coherent picture in which the simulated "hedgehog" molecular configurations directly explain the experimentally observed decrease in dielectric permittivity. This approach establishes a bridge between the microscopic order and the macroscopic dielectric response of the system.

\section{Materials and Methods}
\subsection{Experimental}

\subsubsection{Materials}
The liquid crystal used in this study was 4-Cyano-4'-pentylbiphenyl (5CB) with a purity of >99.5\%, purchased from Sigma-Aldrich. Spherical silica nanoparticles (SiO$_2$) with an average diameter of 25 nm were obtained from US Research Nanomaterials, Inc. The nanoparticles were used as received, without any surface modification.

\subsubsection{Sample Preparation}
Two concentrations of LC+NP composites were prepared: 0.1\% and 1.0\% by weight (wt/wt). Prior to dispersion, the SiO$_2$ nanoparticles were annealed in a vacuum to remove surface contaminants. Subsequently, a weighted amount of nanoparticles was added to the 5CB liquid crystal in its isotropic phase ($T > T_c$). The mixtures were then sonicated for several hours using an ultrasonic bath and mechanically homogenized at an elevated temperature until no visible aggregates could be detected under a polarizing optical microscope. This procedure was designed to ensure a uniform dispersion of nanoparticles within the liquid crystal host.

\subsubsection{Broadband Dielectric Spectroscopy (BDS)}

The dielectric measurements were performed using an Alpha A Dielectric Analyzer (Novocontrol, Germany). The samples were placed in a parallel-plate capacitor made of stainless steel with a diameter of 20 mm and a fixed gap of 0.2 mm. The measurements were conducted over a frequency range of $10^{-1}$ Hz to $10^7$ Hz during a controlled cooling cycle from 350 K to 273 K. The cooling rate was maintained at 1 K/min with a temperature stability of $\pm 0.02$ K. An AC measurement voltage of 1 V (rms) was applied, a standard value in liquid crystal studies, which is sufficiently low to avoid inducing field-alignment in the isotropic phase. The static dielectric permittivity, $\varepsilon_s$, was determined from the real part of the complex permittivity at a frequency of $10^4$ Hz.

\subsection{Computer Simulations}

\subsubsection{Simulation system}
The molecular dynamics (MD) simulations were performed on a system consisting of a single spherical SiO$_2$ nanoparticle with a diameter of 6 nm (3,000 atoms) immersed in a host of 10,000 5CB molecules. The system was enclosed in a cubic simulation box of 16.9 nm x 16.9 nm x 16.9 nm with periodic boundary conditions applied in all three dimensions.

\subsubsection{Simulation Protocol}
All simulations were carried out using the NAMD 2.14 software package. The CHARMM force field family was used to describe the interactions in the system. The 5CB molecules were modeled using the CHARMM Force Field. The SiO$_2$ nanoparticle was described using a dedicated, CHARMM-compatible parameter set developed for silica.

The simulations were run in the isothermal-isobaric (NPT) ensemble. The temperature was maintained using a Langevin thermostat, and the pressure was kept at 1 atm using a Nosé-Hoover Langevin piston barostat. The system was first equilibrated at 320 K for 1 ns, followed by a 3 ns production run at the target temperature of 310 K. A time step of 2 fs was used for all simulations.

\subsection{Data Analysis Formalism}
To quantitatively analyze the pretransitional effects observed in the isotropic phase, the following formalism was applied. The anomalous behavior of the static dielectric constant, $\varepsilon_s(T)$, in the vicinity of the I-N transition can be described by a critical-like relation [18-20,26,27]:
\begin{equation}
    \varepsilon_s(T)=\varepsilon^{*}_s +a\left( T-T^{*}\right)+b\left(T-T^{*}\right)^\phi \quad \text{for}\quad 
	T>T_c
\end{equation}
where $\left(T^{*}, \varepsilon^{*}_s\right)$ defines a point in which the continuous phase transition takes place, $\phi$ is a pseudo-critical exponent and $a, b$ are amplitudes. Other most useful form of Eq.(1) is its first derivative:
\begin{equation}
	\frac{d \varepsilon_s(T)}{d T}=a+b\phi\left(T-T^{*}\right)^{\phi-1}.
\end{equation}

The pretransitional anomaly can also be described from a thermodynamic perspective using the concept of Fröhlich entropy, $S_F$. The change in the system's entropy induced by an electric field $E$ is proportional to the first derivative of the dielectric constant [2,16,17]:
\begin{equation}
	S_F(T,E)=S_0(T)+\frac{1}{2}\frac{d \varepsilon_s}{d T} E^2
\end{equation}
where $S_0(T)$ is the entropy at zero field. The Fr\"{o}hlich entropy can be regarded as the change in entropy of the system resulting from the imposition of an electric field $E$ per unit volume.

Combining to Eqs.(2) and (3) and according to the Mean Field theory [1,2,4,5,17,34] one may express the Frh\"{o}lich entropy by critical-like relation:
\begin{equation}
	\frac{\Delta S}{E^2}=\frac{a}{2}+\frac{b\phi}{2}\left(T-T_c\right)^{\phi-1}.
\end{equation}
Let us substitute $\left(T-T_c\right)$ as $z$ and $\frac{\Delta S}{E^2}$ by $y(z)$ and assume that $a<0$, $b>0$, $\phi\neq 1$ and $\phi>0$. Then, the relation $y(z)=a/2+b\phi/2 z^{\phi-1}$ can be examined in terms of its dependence on parameter $\phi$.

This formulation is equivalent to Eq. (2) and provides a thermodynamic basis for analyzing the pretransitional ordering effects observed in the dielectric response.

To analyze the structural and dynamic properties of the simulated system, several key quantities were calculated from the molecular dynamics trajectories.

The global degree of nematic order is quantified by the scalar order parameter, s, defined as the ensemble average of the second Legendre polynomial, $P_2$ [1,2]:
\begin{equation}
	s=\frac{1}{2}\left(3\left\langle cos^2 \theta \right\rangle -1 \right) \equiv 
	\left\langle P_2 \left(  cos^2 \theta \right) \right\rangle 
\end{equation}

where $\theta$ is the angle between the long axis of a liquid crystal molecule and the director, which represents the average orientation axis of the system.

To characterize the local ordering of molecules with respect to the nanoparticle, the radial order parameter, $s_r$, was calculated. It is defined similarly to the global order parameter, but the angle $\theta_r$ is the angle between the long molecular axis and the radial vector originating from the center of the nanoparticle. The average is performed for all molecules within a thin spherical shell at a distance $r$.

The rotational dynamics of the molecules were investigated by calculating the time autocorrelation function of the system's total dipole moment, $M(t)$[2,9,21-24,33]:
\begin{equation}
	\Phi_{\vec{M}} = \frac{ \left\langle \vec{M}(0)\vec{M}(t) \right\rangle}{ \left\langle \vec{M}(0)\vec{M}(0) \right\rangle} \\ ,     \quad \vec{M}=\sum_{i=1}^{N}\vec{\mu}_i
\end{equation}

where $M(t)$ is the vector sum of all molecular dipole moments at time $t$.

\section{Results and Discussion}
\subsection{Dielectric Response and Pretransitional Anomalies}
\begin{figure}
	\centering
	\includegraphics[width=300px]{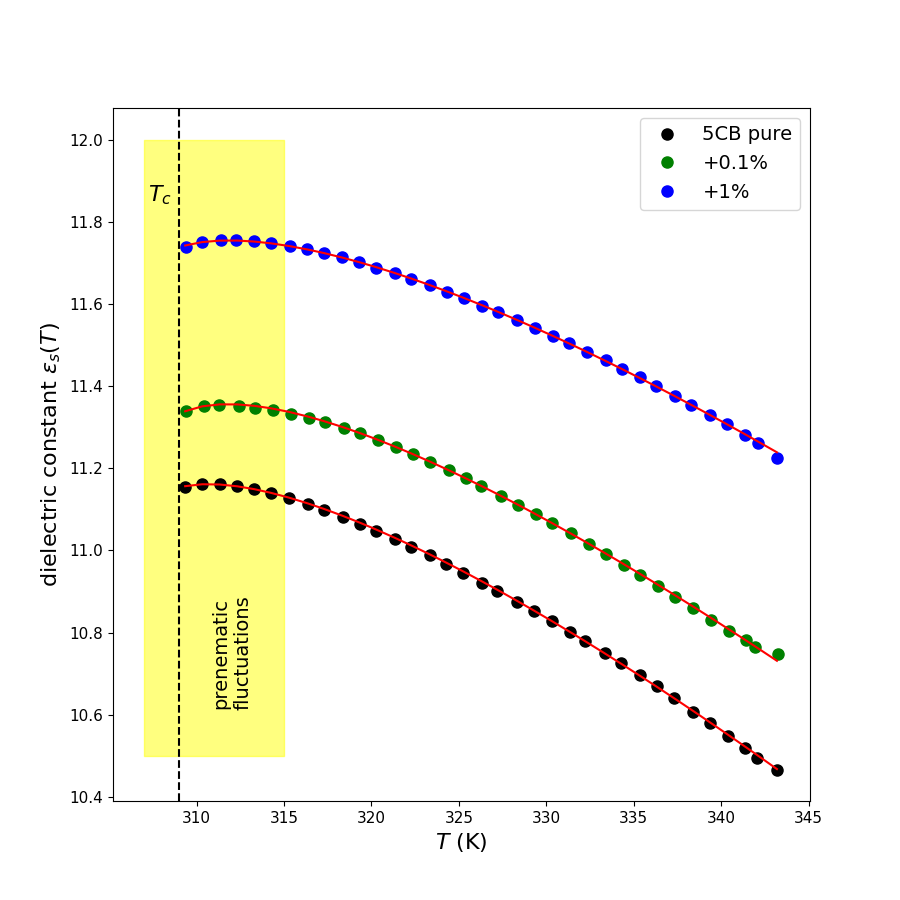}
	\caption{Temperature evolution of dielectric constant ($\varepsilon_s$) in the isotropic phase of pure 5CB liquid crystall and nanoparticles-doped mixtures. In the vicinity of the isotropic-nematic phase transition temperature $T_c$ a strong pretransitional anomaly occurs.}
	\label{fig:enter-label}
\end{figure}

Figure 1 presents the temperature dependence of the static dielectric constant, $\varepsilon_s$, for pure 5CB and for its composites with 0.1\% and 1\% nanoparticle (NP) concentrations. In all samples, a distinct pretransitional anomaly, characterized by a "bending" of the $\varepsilon_s(T)$ curve, is observed in the vicinity of the isotropic-nematic (I-N) phase transition temperature, $T_c$. This well-known effect is attributed to the formation of prenematic fluctuations; within these dynamic domains, the antiparallel arrangement of permanent dipole moments leads to their partial cancellation and a consequent reduction in the measured dielectric constant. A key observation from our data is that while the addition of nanoparticles does not significantly alter the transition temperature itself, it leads to a systematic decrease in the overall magnitude of the dielectric constant across the entire isotropic phase.

To analyze this pretransitional behavior more quantitatively, the first derivative of the dielectric constant with respect to temperature, $\partial \varepsilon_s/ \partial(T)$, is plotted in Figure 2. This derivative is directly proportional to the change in the system's entropy induced by an electric field, known as the Fröhlich entropy ($\Delta S/E^2$), which is plotted on the right axis of Figure 2. The crossover of the derivative from positive to negative values upon heating through $T_c$ corresponds to a change from a more ordered state ($\Delta S/E^2>0$) in the prenematic region to a disordered state ($\Delta S/E^2<0$) in the isotropic phase. This provides a thermodynamic interpretation of the pretransitional ordering. The increased scatter of the data points at temperatures above 335 K, far from the critical region, can be attributed to a lower signal-to-noise ratio.
\begin{figure}
	\centering
	\includegraphics[width=300px]{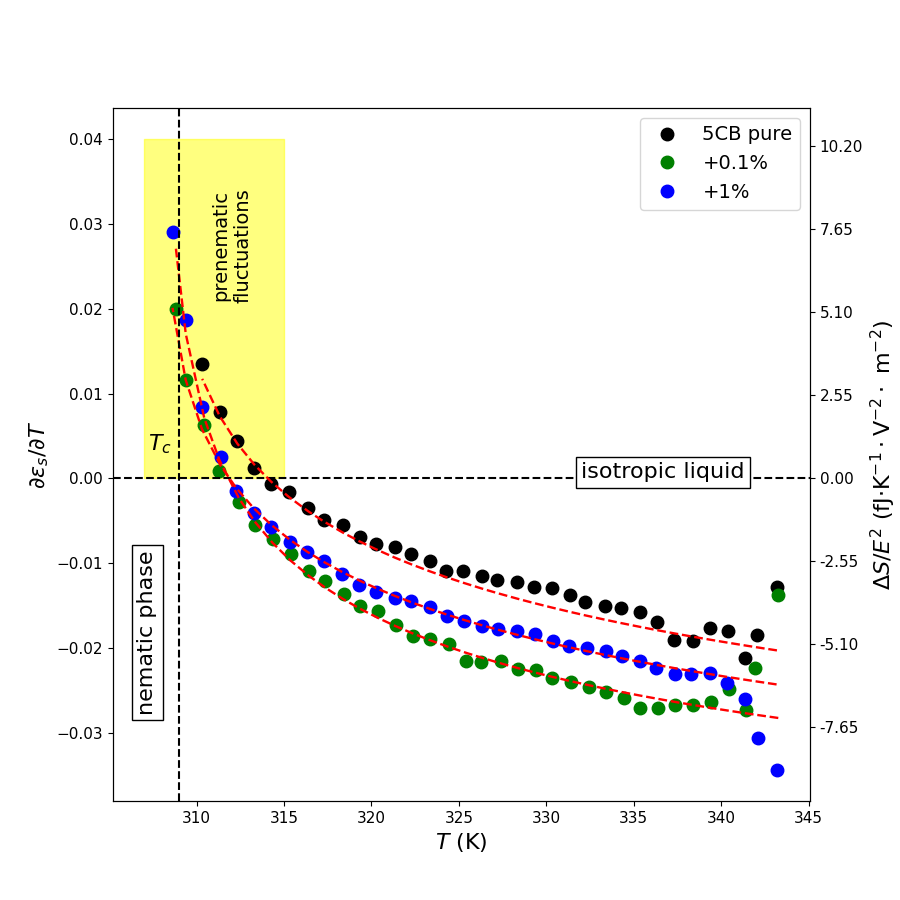}
	\caption{The first derivative of dielectric constant $\partial\varepsilon_s / \partial T$ and Fr\"{o}hlich entropy $\Delta S/E^2$ taken from Eq.(4) for pure and nanoparticles-dopped 5CB. Negative value of $\Delta S/E^2<0$ is related to disordered phase (isotropic), whereas positive $\Delta S/E^2>0$ to ordered one (prenematic).}
	\label{fig:froh}
\end{figure}
To further characterize the nature of the anomaly, the data in the pretransitional region ($T>T_c$) were fitted using the critical-like relation described by Eq. (2) and (4). The dashed red lines in Figure 2 represent the best fits to the experimental data. A crucial outcome of this analysis is that the determined critical exponents, which describe the nature of the singularity, are found to be identical for both the pure and the NP-doped systems within the experimental error. This result provides strong evidence for the universality of the critical phenomena governing the I-N transition, indicating that the fundamental character of the prenematic fluctuations is not perturbed by the presence of the nanoparticles at these concentrations.

However, while the character of the pretransitional effects remains universal, the systematic decrease in the magnitude of $\varepsilon_s$ upon NP doping requires a microscopic explanation. To elucidate the molecular-level origin of this behavior, we subsequently turned to molecular dynamics simulations.

\subsection{Microscopic Origin of the Dielectric Behavior}

To elucidate the molecular-level mechanism behind the observed decrease in dielectric permittivity upon NP doping, we performed molecular dynamics (MD) simulations. The simulation cell, containing a single SiO$_2$ nanoparticle within a 5CB host, is visualized in Figure 3. The results reveal the formation of distinct, highly ordered layers of LC molecules around the nanoparticle. Specifically, a layer of molecules arranged radially with respect to the nanoparticle's surface forms a structure referred to as a "hedgehog".
\begin{figure}
	\centering
	\includegraphics[width=300px]{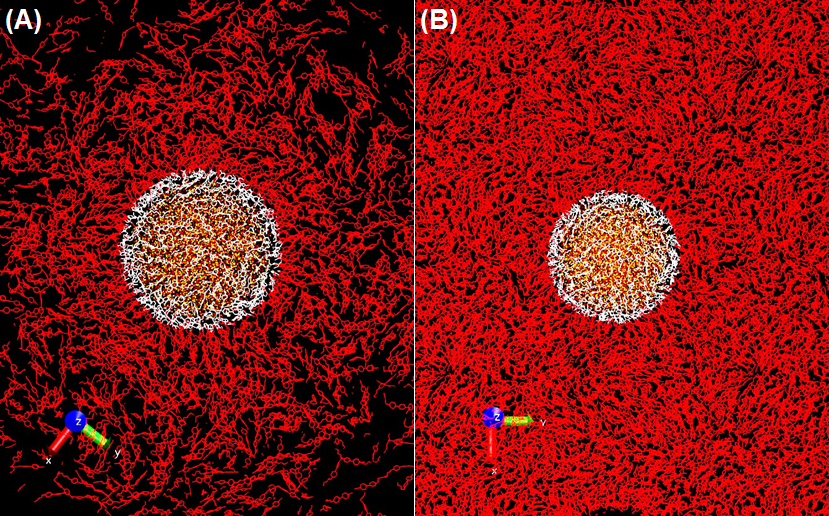}
	\caption{The simulation cell at $T=310$ K, nanoparticle diameter $2r=6$ nm. Molecules in the planar layer are marked in white. In part (A), a clearly visible ordered layer forms a \textit{hedgehog}-like structure. Part (B) takes into account periodic boundary conditions (PBC).}
	\label{fig:enter-label}
\end{figure}
This local ordering was quantified by calculating the radial order parameter, $s_r$, as a function of distance from the center of the nanoparticle, shown in Figure 4. The plot exhibits a sharp negative dip, reaching $s_r \approx -0.38$ , which corresponds to a thin planar layer of molecules directly at the nanoparticle surface. Immediately following this layer, a prominent positive peak with $s_r \approx 0.25$ is observed, which is the quantitative signature of the radially aligned hedgehog structure. Beyond this ordered region ($r>40$ \AA), the order parameter decays to zero, indicating the bulk isotropic phase. These simulations provide clear evidence of a static, NP-induced local order that persists within the globally isotropic phase.
\begin{figure}
	\centering
	\includegraphics[width=200px]{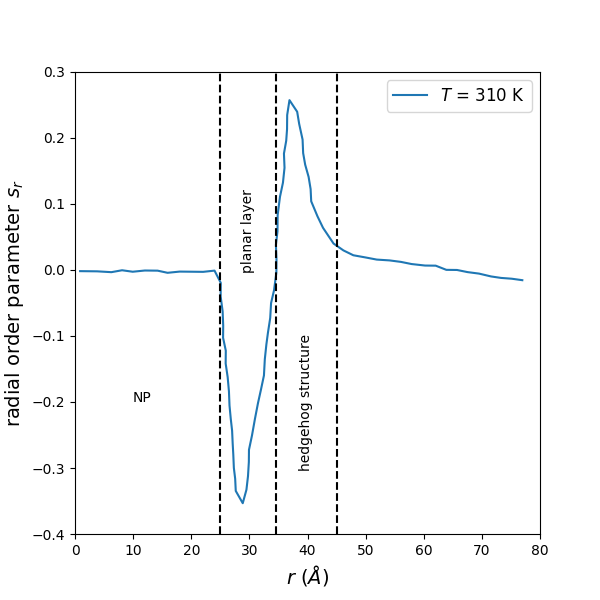}
	\caption{Radial order parameter $s_r$ as a function of distance from the center of the nanoparticle in simulated system based on 5CB at a temperature $T=310$ K.}
	\label{fig:enter-label}
\end{figure}
\subsection{Molecular Arrangement and Dielectric Permittivity}

The simulation results provide a direct microscopic explanation for the experimentally observed decrease in the dielectric constant. The 5CB molecule possesses a strong permanent dipole moment aligned along its long axis. The "hedgehog" structure, characterized by the radial alignment of these molecules around the nanoparticle, therefore imposes an antiparallel arrangement of their dipole moments when considered over the entire nanoparticle surface.

This antiparallel configuration leads to a local cancellation of the dipolar contributions to the permittivity in the immediate vicinity of each nanoparticle. Consequently, each nanoparticle and its ordered shell effectively create a domain with a significantly reduced local dielectric constant. Since the macroscopic measurement averages the dielectric response over the entire sample volume, the presence of these low-permittivity domains results in a lower overall value of the measured $\varepsilon_s(T)$. This effect becomes more pronounced with increasing nanoparticle concentration, which is fully consistent with the experimental data presented in Figure 1.

It is important to note that the nanoparticle, with its induced static shell of paranematic order, acts as a seed that locally mimics the antiparallel dipolar arrangement found in dynamic critical fluctuations. These two phenomena—one static and extrinsic, the other dynamic and intrinsic—coexist, and both contribute to the overall dielectric response of the system.

\subsection{Nematic Order Parameter}
For thermotropic liquid crystals characterized by a phase transition from the isotropic phase to the nematic phase, the phase transition temperature is simultaneously the critical temperature $T_{IM} \equiv T_c$. From a mesoscopic point of view, the orientational order is described by a tensor order parameter $\underline{Q}$. In the case of uniaxial alignment, it takes the form [1-3,6,11,18,20,25,27]
\begin{equation}
	\underline{Q}=s(\vec{n}\otimes\vec{n} -\underline{I}/3)
\end{equation}
where $\vec{n}$ is identified as the local uniaxial order vector (director), $s$ is a uniaxial order parameter measuring the size of fluctuations around the direction of alignment, and $\underline{I}$ is the unit tensor.
In the simplest case, such as the nematic phase, the order parameter $s$ is defined as follows Eq. (5)[1,3,18,20-25].

Figure 5 presents the orientation of dipole moments in the isotropic (a) and nematic (b) phases. Note, that for isotropic liquid the global ordering equals to zero ($s=0$), whereas in nematics is expressed by Eq.(5). Morover, this equation assumes uniform values of the angle $\theta$ for all particles in the system. It is also worth emphasizing that the order parameter $s$ is equivalent to the Legendre polynomial of the second degree $\left\langle P_2\right\rangle$. The nematic phase differs from the isotropic phase by the value of the order parameter $s$, which ranges between $0.3-0.5$ for the former, while it is zero for the isotropic liquid [2,9].

To determine the value of the macroscopic order parameter $s$, from dielectric data without direct measurements of the dielectric anisotropy, the Haller approximation can be used. This empirical equation relates the order parameter to temperature:
\begin{equation}
    s=\frac{\Delta\varepsilon_s}{\varepsilon_\infty - \varepsilon_0}=s_{iso}+\left[1-\frac{T}{T_c^{*}} \right]^{\beta_c}
\end{equation}
where $s_{iso}$ is the order parameter of the isotropic phase (assumed to be zero), $T_c^{*}$ is a hypothetical continuous transition temperature slightly above the actual clearing point $T_c$, and $\beta_c$ is a pseudo-critical exponent. While this equation was formulated for homogeneous systems, it serves as a useful approximation to compare the ordering trends in the pure and nanoparticle-doped systems.

Figure 6 shows the temperature dependence of the order parameter $s$, calculated using Eq. (8) for the experimental data. The results for the pure 5CB are compared with literature data from optical measurements and computer simulations. The most crucial aspect illustrated in the graph is the increase in the system's order with the rise in nanoparticle concentration, particularly for the 0.1\% admixture. This suggests the presence of a strong ordering interaction between the nanoparticles and the liquid crystal molecules, which is consistent with the formation of ordered hedgehog structures observed in the simulations.

\begin{figure}
	\centering
	\includegraphics[width=400px]{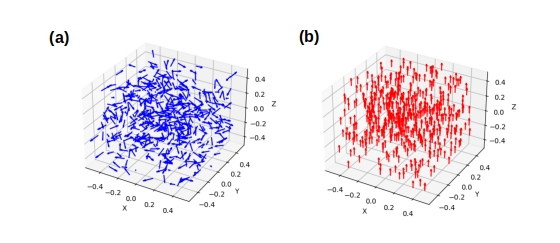}
	\caption{Macroscopic view on dipole moments orientation in (a) isotropic ($T>T_c$) and (b) nematic ($T<T_c$) phases. The order parameter for isotropic liquid $s=0$ and for nematic can be expressed by Eq.(5).}
	\label{fig:enter-label}
\end{figure}

\begin{figure}
	\centering
	\includegraphics[width=1\linewidth]{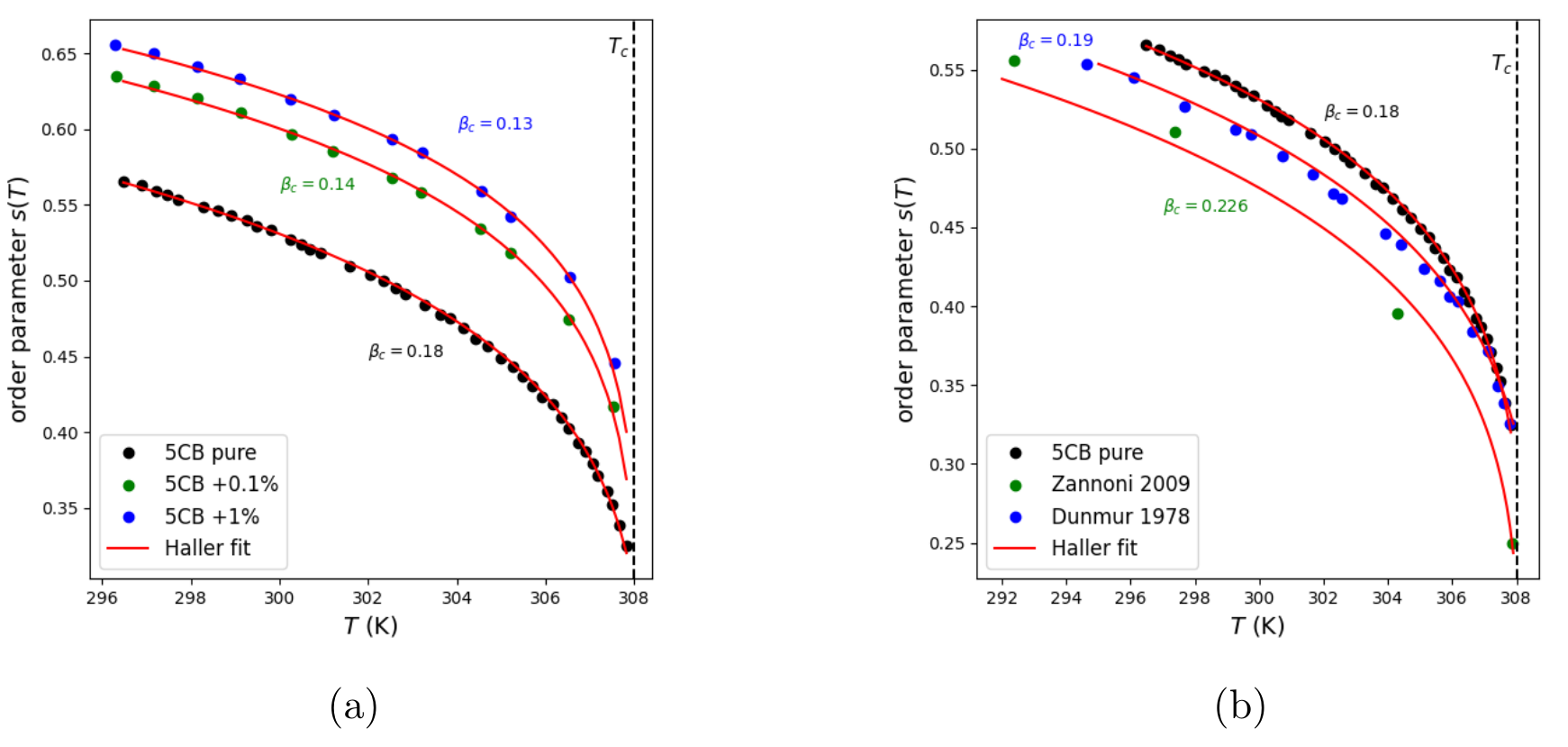}
	\caption{(a) Order parameter $s(T)$ calculated from Eq.(8) for experimental dielectric measurements in nematic phase for various systems based on 5CB liquid crystal. (b) The comparison of the order parameter $s(T)$ calculated from dielectric data, optical measurements [18-20] and computer simulations [2,9,21-24] for nematic 5CB. Red line denotes Haller function (Eq.(8)) fit with given pseudo-critical exponents.}
	\label{fig:obrazki}

\end{figure}
\section{Conclusions}
In this study, we employed an integrated approach, combining broadband dielectric spectroscopy (BDS) and molecular dynamics (MD) simulations, to study the influence of nanoparticle inclusions on the pretransitional phenomena in a nematic liquid crystal. Our findings provide a coherent, multi-scale explanation for the observed dielectric behavior of these composite systems.

The main conclusion of this work is that the experimentally observed decrease in the static dielectric permittivity upon NP doping is a direct consequence of nanoparticle-induced topological defects.  MD simulations explicitly demonstrate that nanoparticles act as "seeds" for local ordering, forcing the surrounding liquid crystal molecules into a radial "hedgehog" configuration.  This static, ordered structure imposes an antiparallel arrangement of molecular dipoles, leading to their local cancellation and a significant reduction in the local dielectric constant.  The macroscopic decrease in permittivity is therefore an average effect resulting from the presence of these low-permittivity defect domains within the sample. 

Furthermore, our analysis of the pretransitional anomaly in the isotropic phase reveals that the associated critical exponent is independent of the nanoparticle concentration.  This confirms the principle of universality for the isotropic-nematic phase transition, indicating that the fundamental character of the intrinsic, dynamic critical fluctuations is not perturbed by the presence of the extrinsic, static defects induced by the nanoparticles. 

This work successfully fills the gap between macroscopic observations and microscopic mechanisms. It highlights an  interplay between static, defect-induced order and dynamic, critical fluctuations, underscoring the power of a combined experimental and computational methodology to unravel the complex physics of nanoparticle-hosted soft matter systems.

\section{Appendix: Theoretical Framework}
\subsection{Dielectric Properties of Anisotropic Media}
The electric permittivity is described as a second-rank tensor for anisotropic media such as liquid crystals, where the electric displacement field $D$ is related to the electric field $E$ by $D=\varepsilon \varepsilon_0E$. For uniaxial phases like nematics, with the director $\vec{n}$ aligned along the z-axis, this tensor has two principal components: $\varepsilon_\parallel$ (parallel to $\vec{n}$) and $\varepsilon_\perp$ (perpendicular to $\vec{n}$). The dielectric anisotropy is then defined as $\Delta\varepsilon_s=\varepsilon_\parallel - \varepsilon_\perp$ [1,2,16,17].

According to the Maier-Meier approach, for molecules with a permanent dipole moment $\mu$ at an angle $\theta$ with the long molecular axis, the components of the squared dipole moment are related to the scalar order parameter $s$ by [1,2,16]
\begin{subequations}
	\begin{eqnarray}
		\langle \mu_\parallel^2\rangle=\frac{1}{3}\mu^2\left[1-(3\cos^2\theta-1)s\right]\\
		\langle \mu_\perp^2\rangle=\frac{1}{3}\mu^2\left[1+\frac{1}{2}(3\cos^2\theta-1)s\right]\\
	\end{eqnarray}
\end{subequations}
\subsection{Model of Nanoparticle-Liquid Crystal Interactions}
To consider the influence of nanoparticles on pretransitional effects, we use a simple model of a binary system. We assume the nanoparticles are homogeneously dispersed in the liquid crystal medium and that the coupling is sufficiently weak to not introduce topological defects in this simplified model. The nanoparticles act as seeds for paranematic (weakly ordered) clusters. The free energy density of such a binary system can be described as [11,18-20,25-27]:
\begin{equation}
	f=(1-p_{cl})f_{LC}+p_{cl}f_{cl}+p_{cl}(1-p_{cl})f_{int}
\end{equation}
where $f_{LC}=A_0(T-T^{*})s^2-Bs^3+Cs^4$, $f_{cl}=a_0(p^{*}-p_{cl})s^2_{cl}-bs^3_{cl}+cs^4_{cl}$ and $f_{int}=Ws_{cl}s$.
Here, $s$ and $s_{cl}$ determine the nematic and paranematic order, respectively. The parameters $A_0, a_0, B, b, C, c, T^{*}, p^{*}$ are material dependent quantities, and 
$p_{cl}=pv_{cl}/v \sim p(r_{cl}/r)^3$. The $f_{cl}$ models the paranematic clusters as effectively lyotropic liquid crystal molecules. The interaction term $f_{int}$ models the coupling between the liquid crystal molecules and paranematic clusters with an interaction strength constant $W>0$. In the situation where $s_{cl}>0$, the liquid crystal component experiences an external field coupling term $f_{int}=-ws$. This can potentially cause a shift in the transition temperature $T_c$ towards higher values. In real systems, however, 
$f_{int}$ is very strongly spatially dependent. It does not have a globally uniform effect of the liquid crystal component. Therefore, in a diluted regime we expect that the phase transition behavior, specifically the value of $T_c$, will still be determined by the liquid crystal component of the binary system. 
\section*{Acknowledgments}
A.I. was supported by the Slovenian Research Agency (ARRS) through grant numbers P2-0232, J2-4447, J3-3066 and J3-3074.

\end{document}